\documentclass{aa}  
\usepackage{adjustbox}
\usepackage[rightcaption]{sidecap}
\sidecaptionvpos{figure}{c}
\usepackage[version-1-compatibility]{siunitx} 
\bibpunct{(}{)}{;}{a}{}{,} 
\begin{document}

   \title{An accurate strong lensing model of the Abell 2163 core}

   \subtitle{}

   \author{U. Rescigno\inst{1}\fnmsep\thanks{umberto.rescigno@unimi.it}
          \and
          C. Grillo\inst{1}\fnmsep\inst{2}
          \and
          M. Lombardi\inst{1}
          \and
          P. Rosati\inst{3}\fnmsep\inst{4}
          \and
          G. B. Caminha\inst{5}
          \and\\
          M. Meneghetti\inst{4}\fnmsep\inst{6}
          \and
          A. Mercurio\inst{7}
          \and
          P. Bergamini\inst{8}
          \and
          D. Coe\inst{9}
         }

   \institute{
    Dipartimento di Fisica, Università degli Studi di Milano, via Celoria 16, I-20133 Milano, Italy\\
    \and
    Dark Cosmology Centre, Niels Bohr Institute, University of Copenhagen, Lyngbyvej 2, 2100 Copenhagen, Denmark\\
    \and
    Dipartimento di Fisica e Scienze della Terra, Università degli Studi di Ferrara, Via Saragat 1, I-44122 Ferrara, Italy\\
    \and
    INAF - Osservatorio di Astrofisica e Scienza dello Spazio, via Gobetti 93/3, 40129 Bologna, Italy\\
    \and
    Kapteyn Astronomical Institute, University of Groningen, Postbus 800, 9700 AV Groningen, The Netherlands\\
    \and
    INFN - Sezione di Bologna, viale Berti Pichat 6/2, 40127 Bologna, Italy\\
    \and
    INAF - Osservatorio Astronomico di Capodimonte, Via Moiariello 16, I-80131 Napoli, Italy\\
    \and
    Dipartimento di Fisica e Astronomia “G. Galilei”, Università degli Studi di Padova, Vicolo dell’Osservatorio 3, I-35122, Italy\\
    \and
    Space Telescope Science Institute, 3700 San Martin Drive, Baltimore, MD 21218, USA\\
   }


\abstract
{Abell 2163 at $z \simeq 0.201$ is one of the most massive galaxy clusters known, very likely in a post-merging phase. Data from several observational windows suggest a complex mass structure with interacting subsystems, which makes the reconstruction of a realistic merging scenario very difficult. A missing key element in this sense is unveiling the cluster mass distribution at high resolution. We perform such a reconstruction of the cluster inner total mass through a strong lensing model based on new spectroscopic redshift measurements. 
We use data from the Multi Unit Spectroscopic Explorer (MUSE) on the Very Large Telescope (VLT) to confirm 12 multiple images of 4 sources with redshift values from 1.16 to 2.72. We also discover four new multiple images and identify 29 cluster members and 35 foreground and background sources. The resulting galaxy member and image catalogs are used to build five cluster total mass models. The fiducial model consists of 111 small-scale subhalos plus a diffuse component, which is centered $\sim2\arcsec$ away from the BCG belonging to the east Abell 2163 subcluster.
We confirm that the latter is well represented by a single, large-scale mass component. Its strong elongation towards a second (west) subcluster confirms the existence of a preferential axis, corresponding to the merging direction. From the fiducial model, we extrapolate the cumulative projected total mass profile and measure a value of $M(<300\,\textnormal{kpc}) = 1.43^{+0.07}_{-0.06}\times10^{14}\,\textnormal{M}_{\odot}$, which has a significantly reduced statistical error compared with previous estimates, thanks to the inclusion of the spectroscopic redshifts. Our strong lensing results are very accurate: the model-predicted positions of the multiple images are, on average, only $0\arcsec.15$ away from the observed ones.}


\maketitle

\section{Introduction}

Our understanding of galaxy clusters has significantly improved in recent years, thanks to high-quality datasets from multi-band surveys with the \textit{Hubble Space Telescope} (\textit{HST}) and spectroscopic follow-up programs with ground-based telescopes. The VIsible Multi-Object Spectrograph (VIMOS) and the Multi Unit Spectroscopic Explorer (MUSE), both mounted on the VLT, have been successfully used to study the mass distribution of galaxy clusters. They have also made it possible to characterize some of the most distant galaxies known to date, as in the Reionization Lensing Cluster Survey (RELICS, \citealt{Coe2019}), whose main goal is to discover and study hundreds of galaxies at $z > 6$ to better understand the epoch of reionization (\citealt{Salmon2017, Salmon2018}). In this program, 46 fields were selected among the most massive \textit{Planck} clusters ($M_{500}>{4\times10^{14}\,\textnormal{M}_{\odot}}$), showing exceptional strong-lensing features. Two other relevant examples are the Cluster Lensing And Supernova survey with \textit{Hubble} (CLASH, \citealt{Postman2012}), a 524-orbit \textit{HST} Multi-Cycle Treasury program targeting 25 high-mass clusters, and the \textit{Hubble} Frontier Fields program (HFF, \citealt{Lotz2017}), an initiative aimed to obtain the deepest \textit{HST} and \textit{Spitzer Space Telescope} observations of six clusters and their lensed galaxies. In this work, we use \textit{HST} data products from the RELICS survey, supplemented with MUSE spectroscopy, and derive the total mass distribution in the core of the cluster Abell 2163 (hereafter A2163) via a strong gravitational lensing analysis.

A2163 is one of the richest Abell clusters, with remarkable features, a complex structure and a variety of interacting subsystems manifesting their presence and activity all over the electromagnetic spectrum. As pointed out in \cite{Soucail2012}, the relations among its mass components are not yet well-explored, thus making hard to define a clear picture of the cluster physical state. In particular, the relations between its different components have never been confronted via detailed strong lensing studies based on spectroscopic measurements. With our work we bridge this gap and provide redshift measurements of all multiple images previously detected and of the new ones we have identified.

All A2163 properties (see Sect. \ref{sec:A2163}) suggest that its core is in a non relaxed state and is undergoing some sort of (post-)merging process, which adds complexity to the dynamical properties of this system. To reconstruct a realistic scenario for A2163, it is thus necessary to use all information at our disposal and combine all mass measurements. In this cluster, X-ray, optical and weak lensing studies are all present in the literature, with mass estimates somehow discordant \citep{Okabe2011, Bourdin2011}, while strong lensing mass models rely on photometric redshifts only (\citealt{Cerny2018}, hereafter C18). The present work is intended to complete the results of the other mass diagnostics, providing a strong lensing mass estimate of the core of A2163.

Finally, for massive clusters where the first merging passage has already occurred, a further intriguing research opportunity is offered by deriving a high-resolution mass map, like the one provided here. In fact, it should be possible to test the presence of the dark matter (DM) self-interaction \citep{Spergel2000} by analyzing the displacements of the three mass components of each cluster merging substructures \citep{Markevitch2004, Harvey2015}: the DM halo, the galaxies and the hot gas. The spatial configuration representative of such a DM scenario predicts an X-ray peak located in the barycenter of the two merging systems, due to the gas collisional behavior; then, from here towards two opposite directions, there should be the centers of the two self-interacting DM halos associated with each subcluster and, to follow, the luminous galaxy component, which is collisionless.

The outline of the paper is as follows. In Sect. \ref{sec:A2163}, we introduce A2163 overall, its multi-wavelength characteristics and its subcomponents.
We then release our catalog of all reliable, spectroscopic redshifts in Sect. \ref{sec:Observations&data} (Table \ref{tab:spec_members_cat}), where the cluster \textit{HST} imaging and MUSE spectroscopic observations are described. In Sect. \ref{sec:StrongLensingModeling}, we detail the selection criteria for both cluster members and multiple images and present our strong lens modeling of A2163; here, we also describe the method adopted to derive the total, projected mass profile of its core, which is shown in Sect. \ref{sec:results}, together with other results. Finally, in Sect. \ref{sec:Conclusions}, we compare our work to the literature and sketch our conclusions.

Throughout this paper, we adopt a $\Lambda$CDM cosmology with $\Omega_{M}=0.3$, $\Omega_{\Lambda}=0.7$, and $H_0=70$ km s$^{-1}$ Mpc$^{-1}$, so that, in this cosmology, $1\arcsec$ corresponds to a physical scale of $3.31$ kpc at the cluster redshift ($z_{\mathrm{Lens}}=0.201$). Moreover, all magnitudes are measured in the AB system ($AB:=31.4-2.5\log\left\langle f_{\nu}/\textnormal{nJy}\right\rangle$) and images are oriented north-east, with north at top and east to the left, with angles measured counterclockwise, from the west direction.

\section{Abell 2163}\label{sec:A2163}

Located at $z\simeq0.201$, A2163 is the most massive galaxy cluster of the RELICS Survey, with a mass $M_{500}$ of approximately $1.6\times10^{15}\,\textnormal{M}_{\odot}$, as estimated from the \textit{Planck} collaboration \citep{Planck2015results} and reported in C18.

A2163 is among the most luminous clusters in X-rays ($L_{X[2-10\textnormal{ keV}]} = 6.0\times10^{45}$ erg s$^{-1}$, \citealt{Elbaz1995}), with exceptionally high gas temperatures, varying between $11.5$ and $14.6$ keV \citep{Arnaud1992, Elbaz1995, Markevitch1996, Markevitch2001}. The gas distribution is non-isothermal, with a high temperature gradient in the center and a strong temperature decline in the outer regions \citep{Markevitch1994, Govoni2004, Ota2014}; its generally complex gas distribution shows features similar to those observed in the Bullet cluster (e.g. \citealt{Soucail2012}).

\cite{Nord2009} presented A2163 maps based on the Sunyaev-Zel'dovich effect (SZ) at two frequencies, $150$ and $345$ GHz, from observations with the APEX-SZ bolometer and the LABOCA bolometer camera, respectively. In combination with archival XMM-Newton X-ray data, SZ measurements were used both to model the radial density and temperature distributions of the intra-cluster medium and to obtain the total mass profile and the gas-to-total mass fraction. Under the assumption of hydrostatic equilibrium and spherical symmetry, the calculated total mass was $M(r<100$ kpc$) = 4-6\times10^{13}\,\textnormal{M}_{\odot}$ and the gas-to-total mass ratio (enclosed within $300$ kpc) was $\sim0.10-0.15$.

As first reported in \cite{Herbig&Birkinshaw}, A2163 radio emission reveals one of the most powerful and extended halos ever detected. This structure, centered in the cluster core, has a quite regular shape and an elongation in the E-W direction, similar to the X-ray emission. An elongated, diffuse source, interpreted as a potential radio relic, is also present in the north-east region \citep{Feretti2001}. The coexistence in the same cluster of a central radio halo and a peripheral relic would suggest a common origin for both structures, to be confirmed by detailed investigations about the cluster merger state \citep{Feretti2001}.

\begin{figure}[ht]
\includegraphics[width=\hsize]{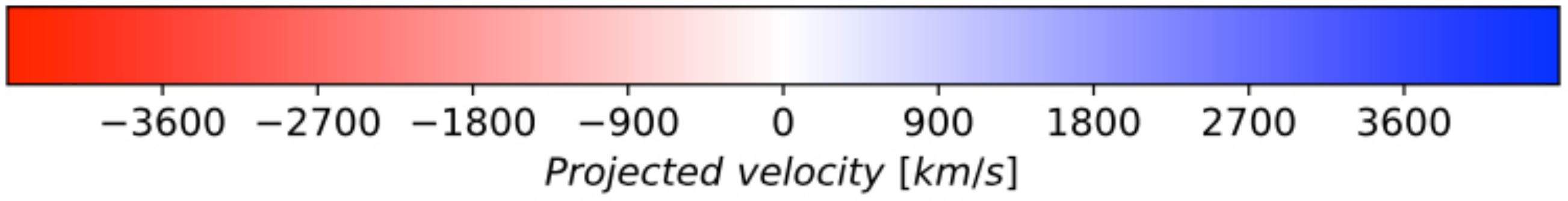}
\includegraphics[width=\hsize]{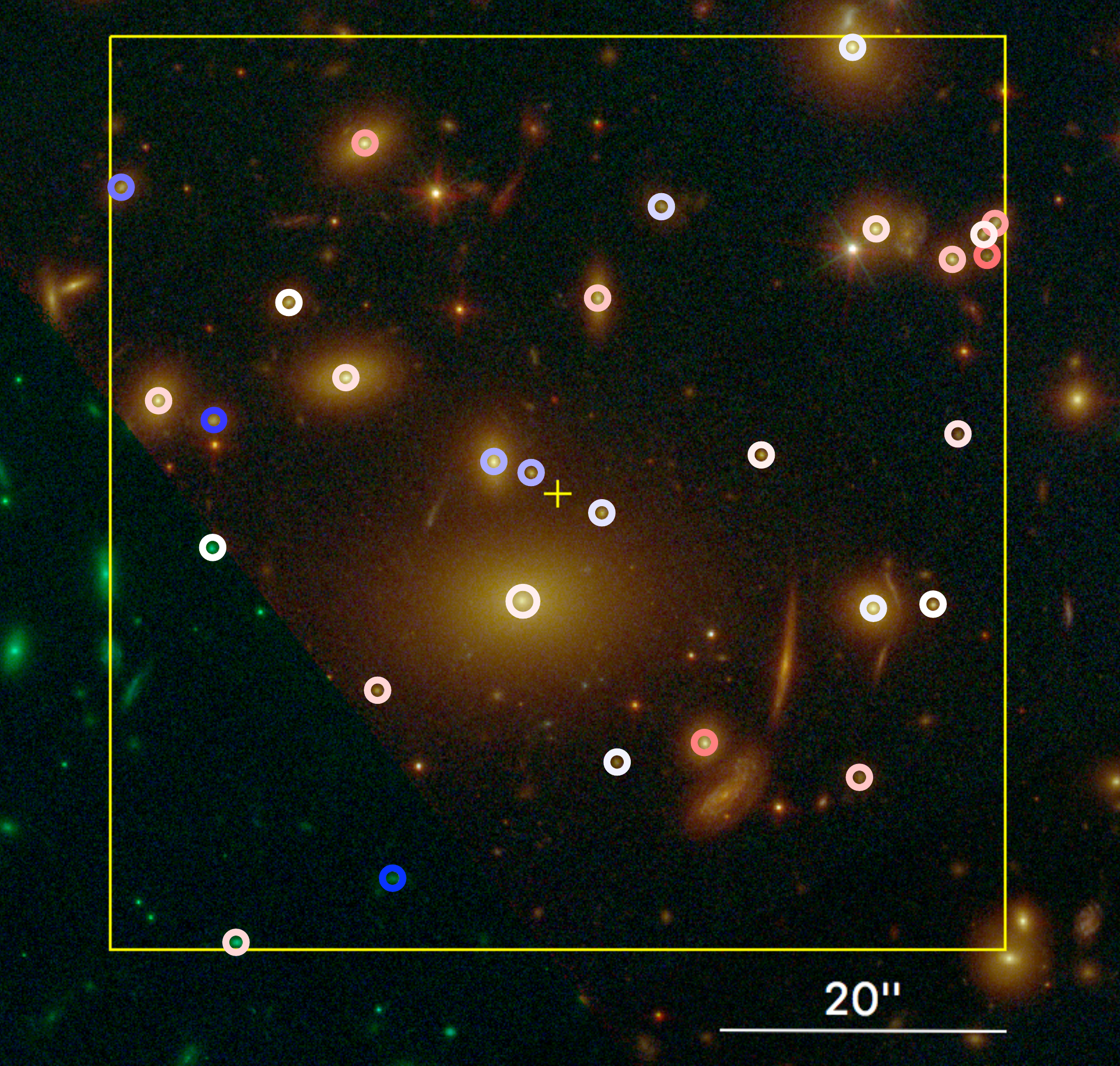}
\caption{\label{fig: MUSE_FoV+spec_cat}Color composite image of Abell 2163 from the \textit{HST} data with the overlaid MUSE pointing (yellow box). The latter is about one arcminute across and is centered on the yellow plus sign position. Circles mark the location of the 29 spectroscopically confirmed galaxy members and they are colored according to the galaxy velocities, relative to the cluster mean redshift. The range of values is shown on the bar on the top.}
\end{figure}

An in-depth optical analysis of A2163 was conducted by \cite{Maurogordato2008} and in this work we will adopt the same names used for its subcomponents (see Fig. 6 of that paper). In the main cluster (A2163-A), galaxies are distributed in two main clumps (A2163-A1, R.A. 16:15:50.9, Decl. $-$06:08:29, in the north-east, and A2163-A2, R.A. 16:15:39.3, Decl. $-$06:09:15, in the south-west), which can be regarded as a pair of colliding structures that will eventually merge into a bigger one. Weak lensing analyses \citep{Okabe2011, Soucail2012} found evidence of a bimodal total mass distribution, with the peak of the hot gas located between the two total mass peaks. Moreover, A2163-A shows a velocity field with a strong gradient, which follows the galaxy distribution and is elongated in the NE/SW direction \citep{Soucail2012}. At larger distances from the main mass clumps, there are several substructures: beyond the north-east radio relic, the most significant one is in the north (A2163-B, R.A. 16:15:48.8, Decl. $-$06:02:21), nearly coincident with a secondary X-ray peak. Studies from spectroscopic and imaging data \citep{Maurogordato2008} confirm that A2163-B is part of the cluster and that is dynamically separated from A2163-A. Nevertheless, between A2163-A and A2163-B, a \textit{bridge} of faint galaxies (along the north-south axis) suggests that the latter is probably infalling into the cluster core \citep{Maurogordato2008}.

In summary, all the results listed above suggest that the cluster core structures, A2163-A1 and A2163-A2, have undergone a recent merger along the elongation direction, while A2163-B is infalling into A2163-A. Although this post-merging scenario is a plausible option, the merging history of A2163 is not yet well-constrained.

\section{Observations and data}\label{sec:Observations&data}

\subsection{\textit{HST} imaging}
We use archival \textit{HST} photometric data in the optical and near-infrared bands ($0.36-1.70$ $\mu$m), taken with seven filters incorporated in the Advanced Camera for Surveys (ACS) and the Wide Field Camera 3 (WFC3). A2163 was observed with the same seven filters adopted in the HFF campaign, for eight ACS and four WFC3/IR orbits, spanning two \textit{HST} pointings as part of two different \textit{HST} programs (a 32-orbit Treasury program, ID GO 12253, cycle 18, P.I. D. Clowe and a 190-orbit program, ID GO 14096, cycle 23, P.I. D. Coe). Further information (e.g., observation dates, exposure times, etc.) is listed in Table 1 of \cite{DePropris2013} and in Table 2 of C18. Figure \ref{fig: MUSE_FoV+spec_cat} shows a color composite image of the cluster core, obtained as the combination of the filters F435W, F606W+F814W and F105W+F125W+F140W+F160W, respectively, for the blue, green and red channels. The reduced \textit{HST} images are publicly available on MAST\footnote{https://archive.stsci.edu/prepds/relics/}.

\subsection{MUSE spectroscopy}
We employed ground-based spectroscopic observations taken with the MUSE instrument \citep{Bacon2012}, mounted on the UT4-Yepun, at the VLT of the European Southern Observatory (ESO).

Specifically, we use spectroscopic data in the north-east inner region of A2163 from the pointing shown in Fig. \ref{fig: MUSE_FoV+spec_cat}. The observations were collected in July 2014 and released in September of the same year, and they were part of one of the instrument commissioning programs (ID 60.A-9100(C), P.I.s MUSE TEAM). A total exposure of four hours was obtained in one single pointing, with an average seeing smaller than $1\arcsec$. The standard data reduction was performed as described in \cite{Grillo2016} and \cite{Caminha2017apr, Caminha2017nov}.

In order to extract spectra and measure redshifts, we consider
the RELICS \textit{HST} ACS+WFC3/IR catalog extending over a total area of $\sim23$ arcmin$^2$ and containing more than 5500 sources,
remove duplications due to segmentation problems, and include 14 additional visually
identified sources. We then extract a total of 230 spectra within circular apertures with radii of $0.6\arcsec$, which belong to the \textit{HST} sources inside the MUSE field of view (FoV); finally, we measure the corresponding redshifts through the software \textit{EZ} \citep{Garilli2010}. Each measurement is tagged with a Quality Flag (QF), which quantifies its reliability, as detailed in \cite{Balestra2016} and \cite{Caminha2016}: \emph{insecure} (QF = 1), \emph{likely} (QF = 2), \emph{secure} (QF = 3), \emph{based on a single emission line} (QF = 9). Redshift estimates with QF = 9 are also considered reliable, since the MUSE spectral resolution makes it possible to identify the shape of narrow emission lines (e.g., Ly$\alpha$) and to distinguish fine-structure doublets (e.g., {[}OII{]}). We obtain a subsample of 64 sources with spectra that have QF $\geq2$: 18 stars, 29 galaxies at $z\simeq0.20$ and 17 (background) galaxies with $z>0.33$ (with 2 high-redshift sources at $z\simeq4.58$ and $z\simeq4.99$). The set of 29 galaxies listed in Table \ref{tab:spec_members_cat} satisfies the cluster membership criteria described in Sect. \ref{subsec:Cluster-members}, they thus represent our sample of spectroscopically confirmed cluster members that we will use in the lensing analysis. The remaining (foreground and background) sources are listed in the table of Appendix \ref{Appendix_A}.

\begin{figure*}[ht]
\includegraphics[width=\hsize]{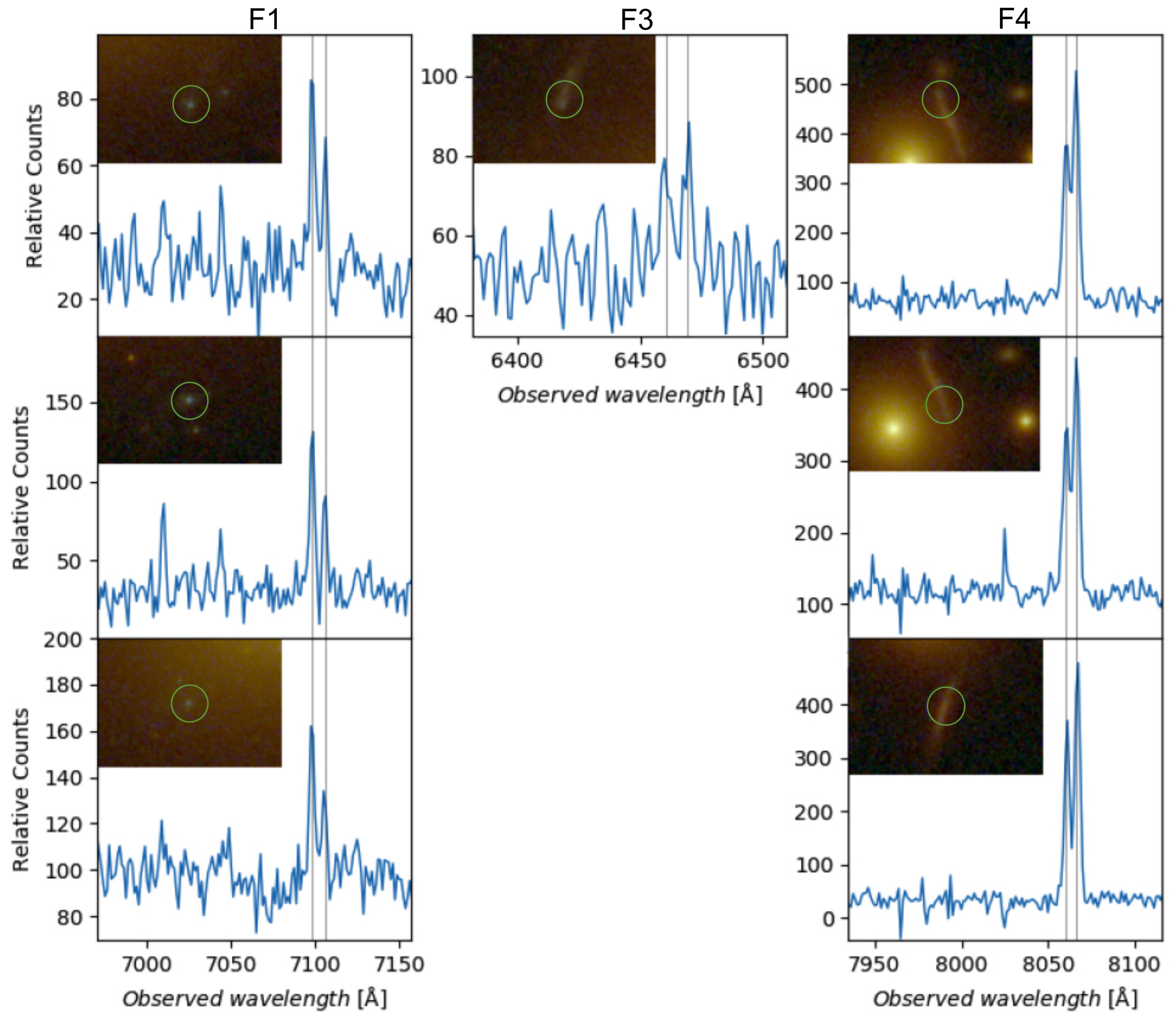}\caption{\label{fig:image spectra}\textit{HST} image cutouts with the circular aperture (green circle) we use to extract the image spectra and the relevant spectral details for some multiple images in A2163 core. The panels of each column refer to the same family and (from the top to the bottom) to the images a, b and c, respectively. QFs for all multiple images are reported in Table \ref{tab:images_cat}. The spectral line doublets (in gray) are [C III] for the first two columns and [O II] for the last one.}
\end{figure*}

With regard to multiple images, we have measured the spectroscopic redshifts of four families, labelled as F1, F2, F3, F4, following the notation in C18. Contrary to the other families, each having three images with at least one secure redshift (QF = 3), F2 has no image with QF > 1. In Fig.~\ref{fig:image spectra} we illustrate, in each panel, one of the reference spectral lines used to estimate the image redshift and a \textit{HST} snapshot centered on each image. The distribution of all multiple images is reported in Fig. \ref{fig:Image_distribution} and more details about them are given in Sect. \ref{images_section}.

In this work, we provide the first spectroscopic confirmation of eight multiple images, the only secure ones known to date for A2163. In fact, in previous studies (e.g., C18), the possible lensed sources in the cluster core were identified only on the basis of their photometric information. There, the photometric redshifts of some multiple images have different values, but the redshift values of each family, optimized in the strong lensing models, are all consistent (within the errors) with ours (see Table 13 in C18).

\begin{table}[ht]
    \caption{Catalog of galaxy cluster members, with \textit{HST} IDs, celestial coordinates, and spectroscopic redshifts with QFs.}             
    \label{tab:spec_members_cat}      
    \centering                        
    \begin{tabular}{c c c c c}        
    \hline\hline                      
    ID & R.A. [Deg] & Decl. [Deg] & $z_{\mathrm{sp}}$ & QF \\    
    \hline                            
   3355	          & 	243.94490	          & 	$-$6.13817	          & 	0.1918	          & 	2	\\
   4431	          & 	243.95039	          & 	$-$6.14759	          & 	0.1930	          & 	3	\\
   3206	          & 	243.95701	          & 	$-$6.13600	          & 	0.1949	          & 	3	\\
   3623	          & 	243.94474	          & 	$-$6.13755	          & 	0.1953	          & 	3	\\
   3461	          & 	243.94558	          & 	$-$6.13824	          & 	0.1973	          & 	3	\\
   3518	          & 	243.95247	          & 	$-$6.13899	          & 	0.1979	          & 	3	\\
   4659	          & 	243.94738	          & 	$-$6.14825	          & 	0.1980	          & 	2	\\
   3820	          & 	243.95677	          & 	$-$6.14657	          & 	0.1990	          & 	3	\\
   3760	          & 	243.96102	          & 	$-$6.14097	          & 	0.1990	          & 	3	\\
   4825	          & 	243.95950	          & 	$-$6.15144	          & 	0.1992	          & 	2	\\
   3686	          & 	243.95737	          & 	$-$6.14052	          & 	0.1998	          & 	3	\\
   3397	          & 	243.94707	          & 	$-$6.13766	          & 	0.1999	          & 	3	\\
   3824	          & 	243.94546	          & 	$-$6.14163	          & 	0.2001	          & 	3	\\
   3829	          & 	243.94929	          & 	$-$6.14202	          & 	0.2008	          & 	3	\\
   3822	          & 	243.95394	          & 	$-$6.14484	          & 	0.2008	          & 	3	\\
   3480	          & 	243.94497	          & 	$-$6.13775	          & 	0.2009	          & 	3	\\
   4051	          & 	243.95998	          & 	$-$6.14382	          & 	0.2017	          & 	3	\\
   4101	          & 	243.94595	          & 	$-$6.14492	          & 	0.2023	          & 	3	\\
   3570	          & 	243.95849	          & 	$-$6.13906	          & 	0.2024	          & 	3	\\
   4544	          & 	243.95211	          & 	$-$6.14796	          & 	0.2029	          & 	2	\\
   2878	          & 	243.94753	          & 	$-$6.13413	          & 	0.2031	          & 	3	\\
   4104	          & 	243.94712	          & 	$-$6.14498	          & 	0.2032	          & 	3	\\
   3821	          & 	243.95240	          & 	$-$6.14314	          & 	0.2039	          & 	3	\\
   3358	          & 	243.95123	          & 	$-$6.13721	          & 	0.2049	          & 	3	\\
   4504	          & 	243.95377	          & 	$-$6.14237	          & 	0.2079	          & 	3	\\
   3933	          & 	243.95450	          & 	$-$6.14216	          & 	0.2079	          & 	3	\\
   3543	          & 	243.96174	          & 	$-$6.13685	          & 	0.2120	          & 	3	\\
   4461	          & 	243.95995	          & 	$-$6.14135	          & 	0.2160	          & 	3	\\
   4890	          & 	243.95647	          & 	$-$6.15021	          & 	0.2191	          & 	3	\\ 
    \hline                                   
    \end{tabular}
    \end{table}

\section{Strong lensing modeling}\label{sec:StrongLensingModeling}

\subsection{Cluster members\label{subsec:Cluster-members}}
Here, we select the galaxy members to include in the mass model of the cluster based on the galaxy spectroscopic and photometric information.

Among the sources with a reliable redshift estimate (i.e., QF $\geq2$), we identify 29 galaxies as being part of A2163 (see Table \ref{tab:spec_members_cat}): they have a redshift distribution which can be fit with a gaussian distribution with mean and standard deviation values of $\bar{z}\simeq0.201$ and $\sigma_{z}\simeq0.006$, respectively. 
This corresponds to a cluster velocity dispersion of $\sigma_{v} \sim1450$~km~s$^{-1}$. The galaxy velocities (relative to the cluster mean redshift) are shown in Fig. \ref{fig: MUSE_FoV+spec_cat} and they are derived as in \citet{Harrison1974}, namely taking into account the main motions that contribute to determining the observed redshift values.

\begin{figure}[ht]
\centering
\includegraphics[width=\hsize]{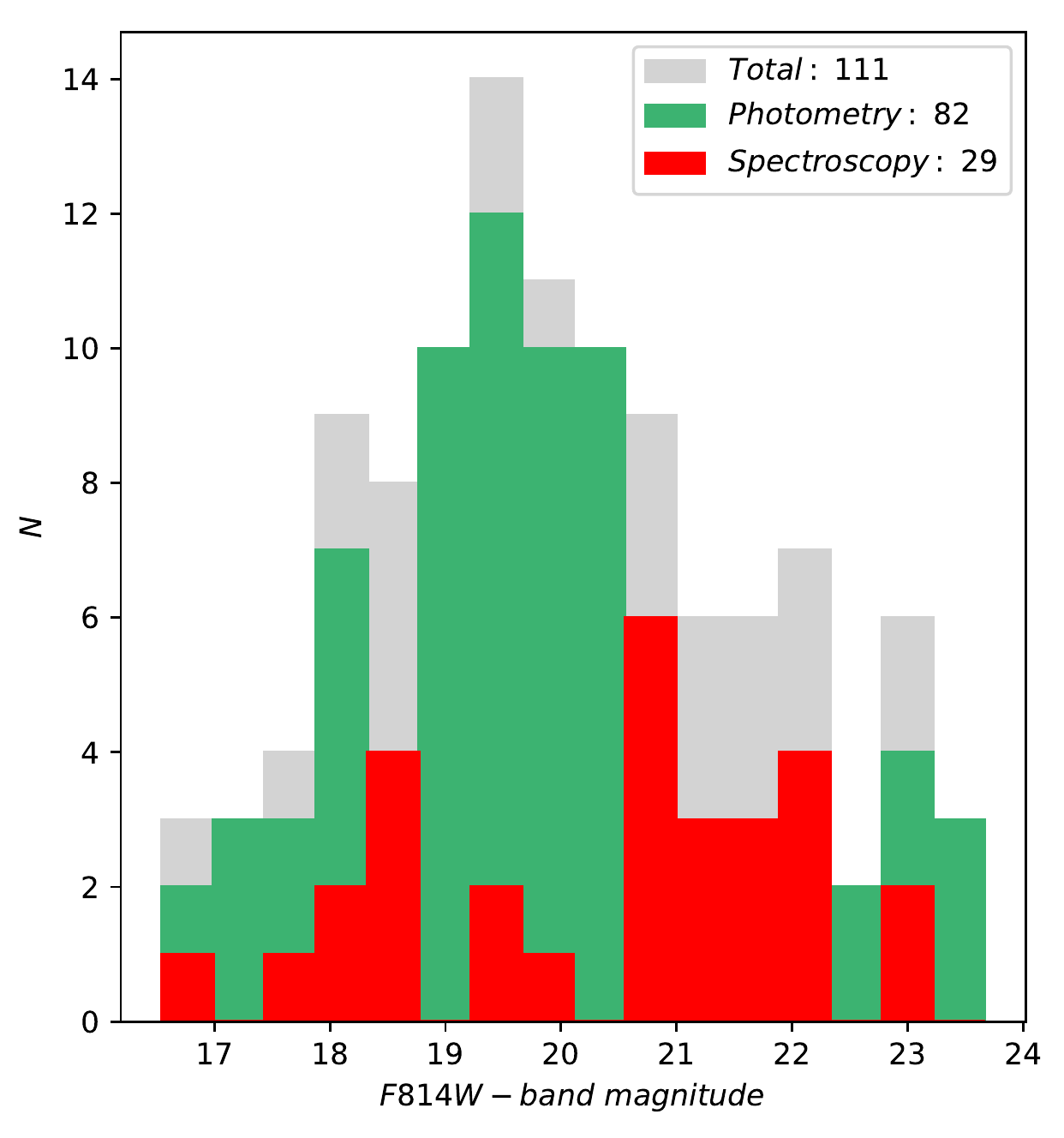}\caption{\label{fig:Probability_cut}Magnitude distribution of cluster members, having $m_{F814W}<24$ and a member probability $P > 95\%$; the whole sample is shown as a gray histogram, while spectroscopically confirmed galaxies (29) and photometrically selected members (82) are in red and green, respectively.}
\end{figure}

As a complement to the above-mentioned spectroscopic catalog, we perform a photometric selection of galaxy members, referring to a novel method based on the extreme deconvolution (\citealt{Bovy2011}) of galaxy color distribution. We define a six-dimensional color space, within which the 7-band galaxy dataset of the whole \textit{HST} catalog is analyzed. We perform a deconvolution of the color distribution of two populations of galaxies, namely the cluster members and the field galaxies, respectively inside and outside the redshift range used to set the membership criterion (i.e., $z$ between 0.19 and 0.22). The deconvolution, which takes into account both the photometric errors and possible incomplete measurements (where one or more bands are missing), produces for each population a representation of the color distribution of member and field galaxies such as a Gaussian Mixture Model (GMM). Since the two populations follow two different GMMs, we can calculate the membership probability for each galaxy on the basis of this probabilistic model. To this purpose, we apply Bayes' theorem, using a $50\%$ prior that the galaxy is a cluster member. Finally, in order to foster cluster galaxy purity over completeness, especially for the reddest sources, we choose a member probability threshold of $P > 95\%$, maximizing the inclusion of the most massive galaxies, which affect the lens model more (see also Sect. 3.3.1 in \citealt{Grillo2015}). In Fig. \ref{fig:Probability_cut}, we report the magnitude distribution of cluster members, obtained with such a choice and with a limiting F814W magnitude of 24 mag. The spectroscopically confirmed foreground and background perturbers are excluded from the lensing models, due to their minor lensing contribution: in the cluster core, we find that background galaxies with $m_\textnormal{F814W} > 19.5$ mag and do not observe foreground galaxies (see the table of Appendix \ref{Appendix_A}). In particular, the background galaxies closest to A2163 ($0.20 < z < 0.41$) are very faint, with $m_\textnormal{F814W} > 24$ mag. \cite{Chirivi2018} found that this choice does not significantly affect the reconstruction of the total projected mass profile. Finally, including these line-of-sight structures would require a multi-plane lensing analysis, which is not yet implemented in the software {\tt lenstool} \citep{Jullo2007}, which we use to model A2163-A1.

With this method, 82 additional photometric members (mostly outside the MUSE FoV) were integrated in the spectroscopic catalog to constitute the final, more complete sample of 111 galaxies used in the lensing analysis.

\subsection{Multiple images}\label{images_section}

The identification of lensed sources is conducted adopting different strategies: we consider all A2163 information available in the literature in light of the indications of our novel data, perform an inspection of both the \textit{HST} image and the MUSE data-cube, and, finally, identify further image candidates, as predicted from a preliminary strong lensing model.

We measure the redshifts of all the lensed systems selected by C18, spectroscopically confirming eight multiple images belonging to three of the four families collected there. We then create a starting model (RUN 1 in Table \ref{tab:Run_cat}) based on a catalog that is a combination of new and previously-detected multiple images: there are 10 in total, and the corresponding MUSE 1D spectra and \textit{HST} snapshots for seven of them are shown in Fig. \ref{fig:image spectra}. Specifically, for the RUN 1 model, we do not use the images of family F2 of C18 and include families F3 and F4. Finally, F1 images were selected based on the following considerations. We identify three new candidates, visually found in the \textit{HST} data. They are labelled as 1c, 2c and 3d, and they lie in the proximity of the BCG, at an approximate distance of five arcseconds ($\sim17$ kpc). Since their photometry is contaminated by that of the BCG, only image 1c was classified as likely (QF = 2). As a result, in our RUN 1 model, we include the new image (1c) and do not consider 2c and 3d (QF < 2).

We then recover a further image (4d), also for family F4: this is a very interesting multiple image system generated by the combined effect of the cluster and a cluster member's gravitational potentials, a situation which is not rare in dense galaxy cluster environments (see, e.g., \citealt{Grillo2014, Parry2016, Caminha2017apr}; Meneghetti et al., in prep.). The galaxy acting as a strong lens is a bright elliptical cluster member, around which a background source is distorted into some arclets (4a, 4b, 4c) and a more compact image (4d). 
A detailed analysis of the system F4 can be found in a separate paper (Bergamini et al., in prep.), where several galaxy-galaxy strong lensing systems in different galaxy clusters are studied.

The image 4d is not visible in the \textit{HST} images, but it can be identified in the subtracted MUSE data-cube. In detail, through the MUSE Python Data Analysis Framework (MPDAF, proceedings of ADASS XXVI, 2016), we sum spatial pixels of the cube over the wavelength interval of the [O II] emission line doublet, which is very prominent in the spectra of the other images of the same family. From this interval, we also eliminate contamination effects by subtracting the background emission taken from two cube slices, below and above the [O II] wavelength range. We also include 4d in the first catalog used in this preliminary stage. Hence, the RUN 1 model consists of three lensed systems with at least one multiple image per system having a secure redshift.

\begin{figure}[ht]
\includegraphics[width=\hsize]{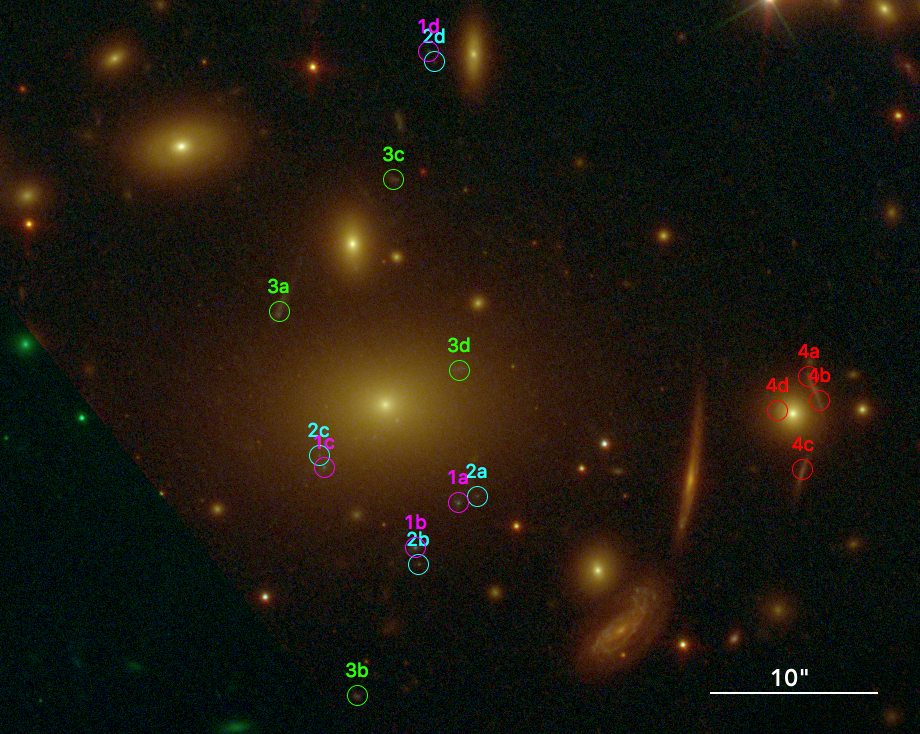}
\caption{\label{fig:Image_distribution}Color composite image illustrating the distribution of the multiple images included in our fiducial model.}
\end{figure}

The complete set of multiple images has a F814W observed magnitude (when present) range of $m_{F814W}\simeq24-29$ mag and spans a redshift range between 1.16 and 2.72, with images of the same family having an equal redshift value (that with the highest QF). They are well-represented by point-like objects (with the exception of the two F4 arcs) and cover the cluster core, targeted by the MUSE observations (Fig. \ref{fig: MUSE_FoV+spec_cat}). The properties of our final catalog of 16 multiple images are reported in Table \ref{tab:images_cat}, and their positions are shown in Fig. \ref{fig:Image_distribution}.

\begin{table}
    \caption{Catalog of multiple images, with ID, celestial coordinates and spectroscopic redshifts with the related QFs; we also report the ID of the reference work in the literature (C18).}             
    \label{tab:images_cat}      
    \centering
    \begin{tabular}{c c c c c c c}        
    \hline\hline                      
    ID & R.A. [Deg] & Decl. [Deg] & $z_{\mathrm{sp}}$ & QF & C18  \\    
    \hline                            
   1a	          & 	243.95273		& 	$-$6.14646 	& 	2.723  & 	3	 & 	1.1 \\
   1b	          & 	243.95345		& 	$-$6.14722 	& 	2.723  & 	3	 & 	1.2	\\
   1c	          & 	243.95497		& 	$-$6.14588 	& 	2.723   & 	2	 & 	 -	\\
   1d	          & 	243.95322 		& 	$-$6.13895 	& 	2.723	& 	2	 & 	1.3	\\
    \hline                                   
   2a	          & 	243.95241		& 	$-$6.14636 	& 	-		& 	0    & 	2.1 \\
   2b	          & 	243.95339		& 	$-$6.14749 	& 	2.264  & 	1    & 	2.2	\\
   2c	          & 	243.95506		& 	$-$6.14568 	& 	2.723	& 	1	 & 	 -	\\
   2d	          & 	243.95312		& 	$-$6.13912 	& 	2.264  & 	1	 & 	2.3	\\
    \hline                                   
   3a	          & 	243.95573		& 	$-$6.14328 	& 	2.389  & 	3	 & 	3.1 \\
   3b	          & 	243.95442		& 	$-$6.14969 	& 	2.389  & 	1	 & 	3.2	\\
   3c	          & 	243.95381		& 	$-$6.14108 	& 	2.389  & 	1	 & 	3.3	\\
   3d	          & 	243.95270		& 	$-$6.14427 	& 	2.389  & 	1	 & 	 -	\\
    \hline                                   
   4a	          & 	243.94685		& 	$-$6.14437 	& 	1.164  & 	3	 & 	4.1 \\
   4b	          & 	243.94668		& 	$-$6.14476 	& 	1.164  & 	3	 & 	4.2	\\
   4c	          & 	243.94695		& 	$-$6.14592 	& 	1.163  & 	3	 & 	4.3	\\  
   4d	          & 	243.94738 		& 	$-$6.14494 	& 	1.164  & 	1    & 	 -	\\
    \hline                                   
    \end{tabular}
    \end{table}

\subsection{Mass components}
We model the total mass distribution in the core of A2163 with two kinds of mass components: a cluster-scale halo, containing mainly DM plus diffuse baryonic matter (i.e., hot gas and stars contributing to the intra-cluster light), and a certain number of small-scale halos, describing each galaxy member and the respective DM substructure around it. As in \cite{Bonamigo2017}, for both components we refer to a dual Pseudo-Isothermal Elliptical (dPIE; \citealt{Eliasdottir2007}) profile, whose parameters can be set according to the features of the specific component to be represented. The dPIE surface mass density, $\Sigma$, is given by
\begin{equation}\label{eq:Theoretical_density_profile}
\begin{cases}
\Sigma\left(x,y\right)=\dfrac{\sigma_{0}^{2}}{2G}\,\dfrac{R_{\mathrm{T}}}{R_{\mathrm{T}}-R_{\mathrm{C}}}\,\left(\dfrac{1}{\sqrt{R_{\mathrm{C}}^{2}+R_{\varepsilon}^{2}}}-\dfrac{1}{\sqrt{R_{\mathrm{T}}^{2}+R_{\varepsilon}^{2}}}\right) \,\,\,, \\\\
R_{\varepsilon}^{2}:=\dfrac{x^{2}}{\left(1+\varepsilon\right)^{2}}+\dfrac{y^{2}}{\left(1-\varepsilon\right)^{2}} \,\,\,, \\\\
\varepsilon:=\dfrac{1-q}{1+q} \,\,\,.
\end{cases}
\end{equation}

Here, $\sigma_{0}$ is the central velocity dispersion, $R_{\mathrm{T}}$ and $R_{\mathrm{C}}$ are the truncation and core radii, respectively; $R_{\varepsilon}$ is the projected radius adjusted to take into account an ellipticity parameter, $\varepsilon$, which is defined through the minor-to-major-axis ratio, $q$. The reference profile has, in general, seven free parameters: the two centroid coordinates, $\sigma_{0}$, $R_{\mathrm{T}}$, $R_{\mathrm{C}}$, $\varepsilon$, and the position angle, $\theta$. For the diffuse component, we only fix the value of the truncation radius to infinity, because strong lensing data cannot constrain this parameter, and keep all the other parameters free. This implies a total of six free parameters. On the other hand, each cluster member is modeled with a spherical dPIE profile, with a vanishing core radius and centered on its luminosity peak. To reduce the number of free parameters associated with the mass contribution of the cluster members, we assume two scaling relations for them:
\begin{equation}\label{eq:scaling_relation_R_T}
R_{\mathrm{T},i}=R_{\mathrm{T,g}}\left(\frac{L_{i}}{L_{\mathrm{g}}}\right)^{0.5},
\end{equation}
\begin{equation}\label{eq:scaling_relation_sig_T}
\sigma_{0,i}=\sigma_{0,\mathrm{g}}\left(\frac{L_{i}}{L_{\mathrm{g}}}\right)^{0.35},
\end{equation}
where $R_{\mathrm{T},i}$, $\sigma_{0,i}$, and $L_{i}$ are, respectively, the values of the truncation radius, central velocity dispersion and F814W luminosity of the $i$-th subhalo; $R_{\mathrm{T,g}}$, $\sigma_{0,\mathrm{g}}$, and $L_{\mathrm{g}}$ are the same quantities for a reference galaxy, which we identify with the BCG ($m_{F814W}\simeq16.55$ mag). We choose these scaling relations because they reproduce the variation of the total mass to light ratio $M\slash{L}$ with luminosity observed in early-type galaxies, known as the tilt of the fundamental plane \citep{Faber1987, Bender1992}. These equations translate into only two free parameters for the cluster small-scale total mass component, $R_{\mathrm{T,g}}$ and $\sigma_{0,\mathrm{g}}$. The cluster member generating the images of family F4 also follows these relations, except for one lens model (see RUN 5 in Sect. \ref{sec:results}).

\subsection{Method description}\label{subsec:method_section}
We infer our final mass model of A2163 through the minimization of the distances between the positions of the observed ($\boldsymbol{\theta}^{\mathrm{obs}}$, with uncertainty $\sigma^{\mathrm{obs}}$) and model-predicted ($\boldsymbol{\theta}^{\mathrm{pred}}$) multiple images. To do that, we use the function 
\begin{equation}\label{eq:chi_squared}
\chi^{2}(\vec{p}):=\sum_{j=1}^{N_{\mathrm{F}}}\sum_{i=1}^{N_{\mathrm{Im},j}}\left(\frac{|{\boldsymbol{\theta}}_{i,j}^{\mathrm{obs}}-{\boldsymbol{\theta}}_{i,j}^{\mathrm{pred}}(\vec{p})|}{\sigma_{i,j}^{\mathrm{obs}}}\right)^{2} ,
\end{equation}
where the subscripts $i$ and $j$ refer, respectively, to the multiple images, in total $N_{\mathrm{Im},j}$ (for the $j$-th family), and the corresponding family, in total $N_{\mathrm{F}}$; $\vec{p}$ is the vector grouping all the model parameters (once the cosmological ones are fixed). As in \cite{Caminha2016}, to quantify the accuracy and the precision of the model, we refer also to the root-mean-square (rms) value of the distances between the observed and model-predicted positions of the multiple images. This quantity is independent of the value of $\sigma^{\mathrm{obs}}$ and is defined as
\begin{equation}\label{eq:rms}
\delta_{\textnormal{rms}}(\vec{p}):=
\sqrt{\sum_{j=1}^{N_{\mathrm{F}}}\sum_{i=1}^{N_{\mathrm{Im},j}}
\frac{\lvert{\boldsymbol{\theta}}_{i,j}^{\mathrm{obs}}-{\boldsymbol{\theta}}_{i,j}^{\mathrm{pred}}(\vec{p})\rvert^{2}}{N}},
\end{equation}
where $N$ is the total number of multiple images.

We adopt the software {\tt lenstool}, which implements the dPIE mass profile described by Eqs. (\ref{eq:Theoretical_density_profile}) and Bayesian Markov Chain Monte Carlo (MCMC) techniques to efficiently explore the posterior distribution of each parameter. We use them to compute the statistical errors and the correlations among the model parameters. Moreover, every run is conducted until convergence, using more than $10^5$ points to sample the posterior probability distribution of the parameters. We choose relatively large uniform priors, setting a conservative range of variation for the parameter values.

In a preliminary run, we tune the uncertainty on the position of the multiple images, to take into account the impact of factors influencing this quantity, such as the clumpiness of the DM distribution in the cluster, the presence of mass interlopers between lensed background sources and the observer and the limitations of parametric mass models \citep{Jullo2010, Host2012}.

Specifically, an initial uncertainty of $0''.10$ is used\footnote{except for the positional error of image 4d, detected only in the MUSE cube; we double it ($0''.20$) because of the different spatial resolution of MUSE data compared to the \textit{HST} images.}, representing about two pixels of the \textit{HST} images (i.e., $0''.06$); then, to include the effects mentioned above, we increase this starting reference value in a second {\tt lenstool} run in order to have a minimum $\chi^2$ value comparable with the number of degrees of freedom (d.o.f.). As previously explained, to model the cluster total mass components, we fix the number of free parameters\footnote{except for two cases, described in in Sect. \ref{sec:results}: RUN 3, where the redshift of one multiple image family is added as a further free parameter and RUN 5, where an additional galaxy-scale halo with two other parameters is considered.} to eight, to which we have to add the two coordinates defining the position of each multiply imaged source. For this reason, the number of d.o.f. varies according to the number of the lensed sources and that of all multiple images, since we use the positions of the latter as constraints.

Finally, the parameter values of the most probable model (i.e., our \textit{fiducial model}) are used to obtain a number of results, such as the projected, total mass profile in the cluster core. 

\begin{figure}[ht]
\centering
\includegraphics[width=\hsize]{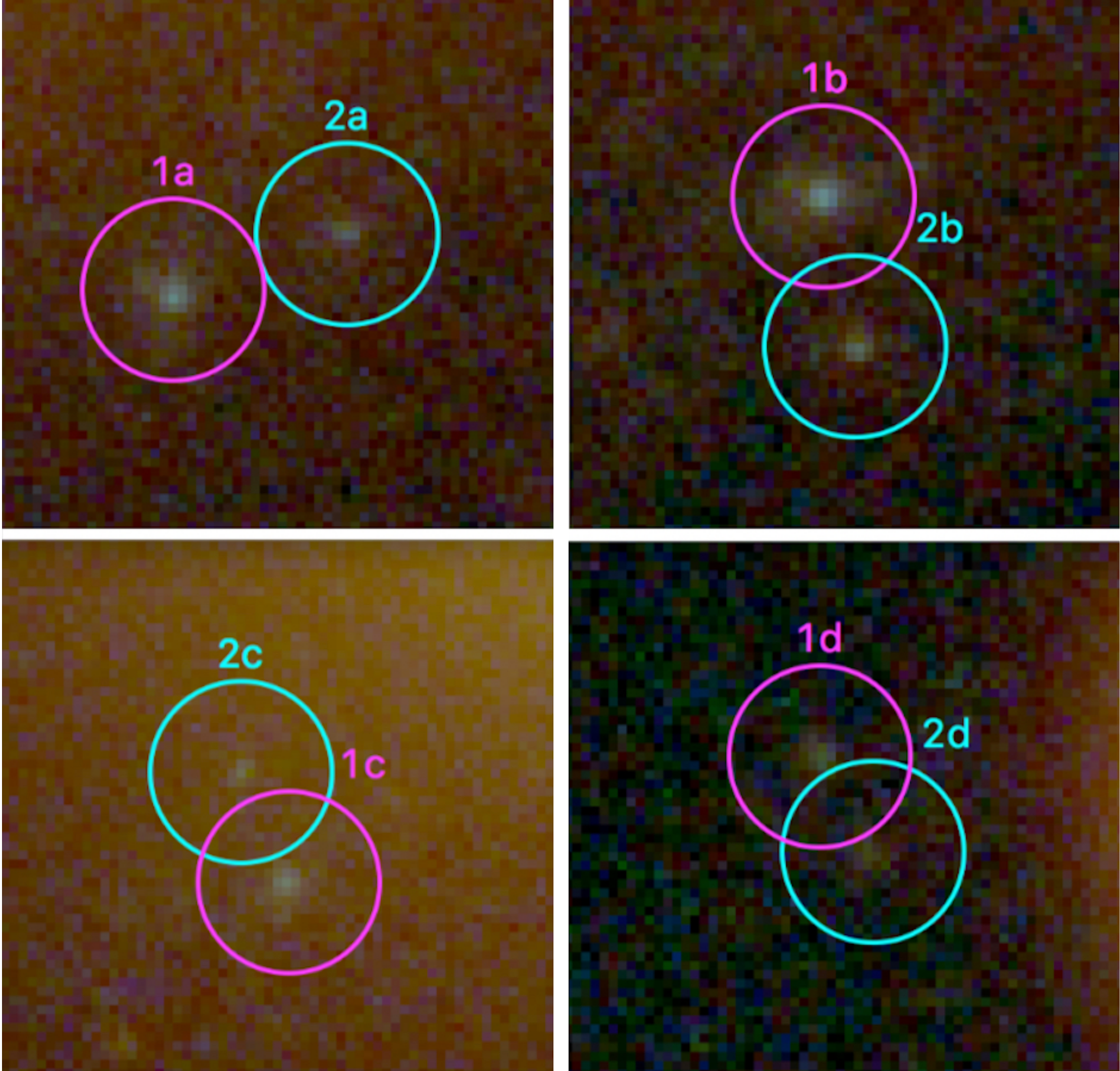}\caption{\label{fig:blob}Collage of \textit{HST} cutouts illustrating the four images of F1 and F2.}
\end{figure}

\begin{figure}[ht]
\centering
\includegraphics[width=\hsize]{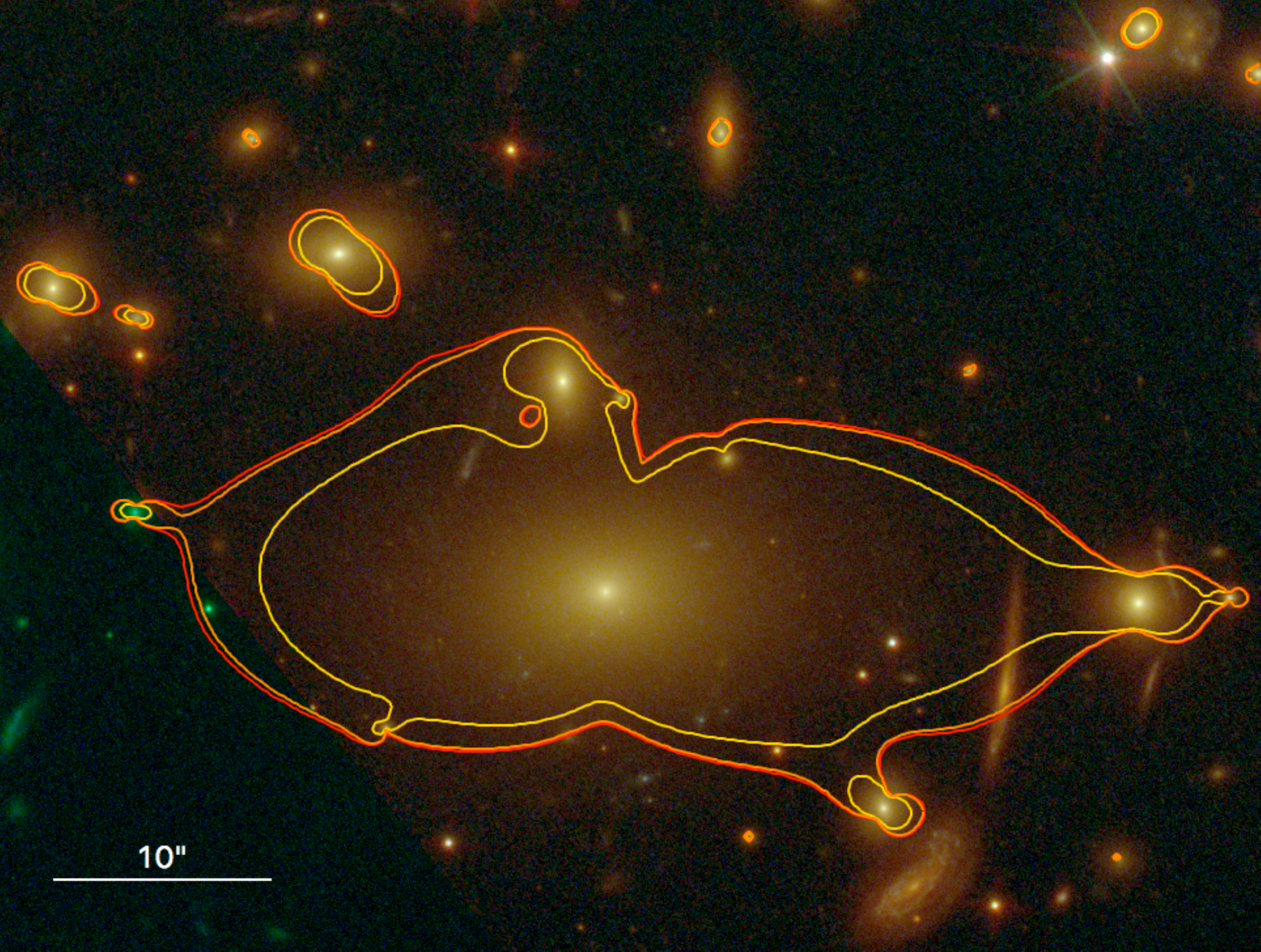}\caption{\label{fig:critical_lines}Critical lines of the fiducial model, for sources at redshifts 1.164 (gold), 2.389 (orange) and 2.723 (red), superposed on a color composite image of A2163-A1.}
\end{figure}

\section{Results and discussion}\label{sec:results}

With the method detailed in Sect. \ref{subsec:method_section}, we explore the results of a collection of runs, reported in Table \ref{tab:Run_cat}. We start from the model presented in Sect. \ref{images_section} (RUN 1), in which we consider a minimal number of multiple-image systems and a reduced version of our final catalog of cluster members (only including cluster galaxies spectroscopically confirmed). We then refine the model choices of each successive run considering the results of the previous one. For example, we include further images, properly selected on the basis of the findings of each preceding run. We now describe the analysis process from the first model to the fiducial one (corresponding to RUN 5). In this model, the adopted multiple images are those given in Table \ref{tab:images_cat}. 

\begin{table}
    \caption{Summary of run characteristics and results. For each run, we report the ID, the images included in the corresponding model, and its relevant statistical quantities: the initial value of the minimum $\chi^2$ (the final one equals the number of d.o.f., column 4) and the rms error, as defined in Sect. \ref{subsec:method_section}.} 
    \label{tab:Run_cat}      
    \centering                        
    \begin{tabular}{c c c c c c}        
    \hline\hline                      
    RUN ID & images & $\chi^2_{in}$ & d.o.f. & $\delta_{\textnormal{rms}}$ \\    
    \hline                            
   1 & {$1a, b, c$}&\\&{$3a, b, c$}&\\&{$4a, b, c, d$} & 52.11 & 6 & $0\arcsec.25$ \\
   2 & {$1a, b, c, d$}&\\&{$3a, b, c, d$}&\\&{$4a, b, c, d$} & 72.34 & 10 & $0\arcsec.26$ \\
   3 & {$1a, b, c, d$}&\\&{$2a, b, c, d$}&\\&{$3a, b, c, d$}&\\&{$4a, b, c, d$} & 76.38 & 15 & $0\arcsec.24$	\\
   4 & {$1a, b, c, d$}&\\&{$2a, b, c, d$}&\\&{$3a, b, c, d$}&\\&{$4a, b, c, d$}  & 76.80 & 16 & $0\arcsec.24$	\\   
   5 & {$1a, b, c, d$}&\\&{$2a, b, c, d$}&\\&{$3a, b, c, d$}&\\&{$4a, b, c, d$}  & 35.06 & 14 & $0\arcsec.15$	\\
    \hline                                   
    \end{tabular}
    \end{table}

For the first model (RUN 1), a final rms offset between the observed and model-predicted positions of the multiple images of $\delta_{\textnormal{rms}}=0\arcsec.25$ is found. This model predicts two further counter images for families 1 and 3 (1d and 3d), angularly very close to those we have visually identified as possible candidates. For this reason, they are included in a second, larger catalog. Thus, the second model (RUN 2) reproduces the multiplicity of all families and consists of three systems of four images each. It leads to a comparable final rms error of $\delta_{\textnormal{rms}}=0\arcsec.26$. Only at this stage, we consider the four images of F2, all of which have a quality flag QF < 2: three come from the literature (C18), and the remaining one (2d) was identified by us in the \textit{HST} data. The addition of F2 to the previous image catalog is justified by considering that the corresponding background source seems connected to that of F1, as illustrated in Fig. \ref{fig:blob}. We find that the final rms error reduces when the F2 source redshift, $z_{\textnormal{F2}}$, is free to vary (RUN 3). However, a similar rms value is recovered when we consider an additional run, identical to RUN 3, except for $z_{\textnormal{F2}}$, which we fix to the redshift value of F1. Moreover, the model-predicted redshift, $z_{\textnormal{F2}}=2.666\pm{0.174}$, has a value consistent with that measured for family F1, with a difference of $\Delta z\sim 0.06$. For the last two runs (RUN 4 and RUN 5), we consider an extended cluster member catalog, including both spectroscopic and photometric members (as detailed in Sect. \ref{subsec:Cluster-members}). Compared to RUN 3, in RUN 4 (where $z_{\textnormal{F2}}\equiv z_{\textnormal{F1}}$), the logarithm of the evidence increases, $\delta_{\textnormal{rms}}$ does not improve significantly, and the number of free parameters is smaller. The BCG velocity dispersion value is $\sigma_{0,\mathrm{g}} \simeq 400$ km s$^{-1}$. We seek to explain this quite high value, by analyzing how the scaling relations (\ref{eq:scaling_relation_R_T})-(\ref{eq:scaling_relation_sig_T}) work in practice. We find that the values of the parameters of the scaling relations are driven by the galaxy around which the family F4 is observed, consequently the high velocity dispersion value of the BCG derives from the assumption that the same relations hold for all the cluster members. A comparison between our value of the BCG velocity dispersion with that derived by other strong lensing studies is not possible here, due to the the lack of such information in the literature. Based on these considerations, in RUN 5, we free the galaxy contributing to the formation of F4 arcs from the scaling relations and model it with a spherical isothermal mass profile, with two additional free parameters, the central velocity dispersion, $\sigma_{0,\mathrm{g2}}$, and the truncation radius, $R_{\mathrm{T,g2}}$. RUN 5 provides the best results, thus we refer to it to illustrate our findings.

\begin{figure}
\includegraphics[width=\hsize]{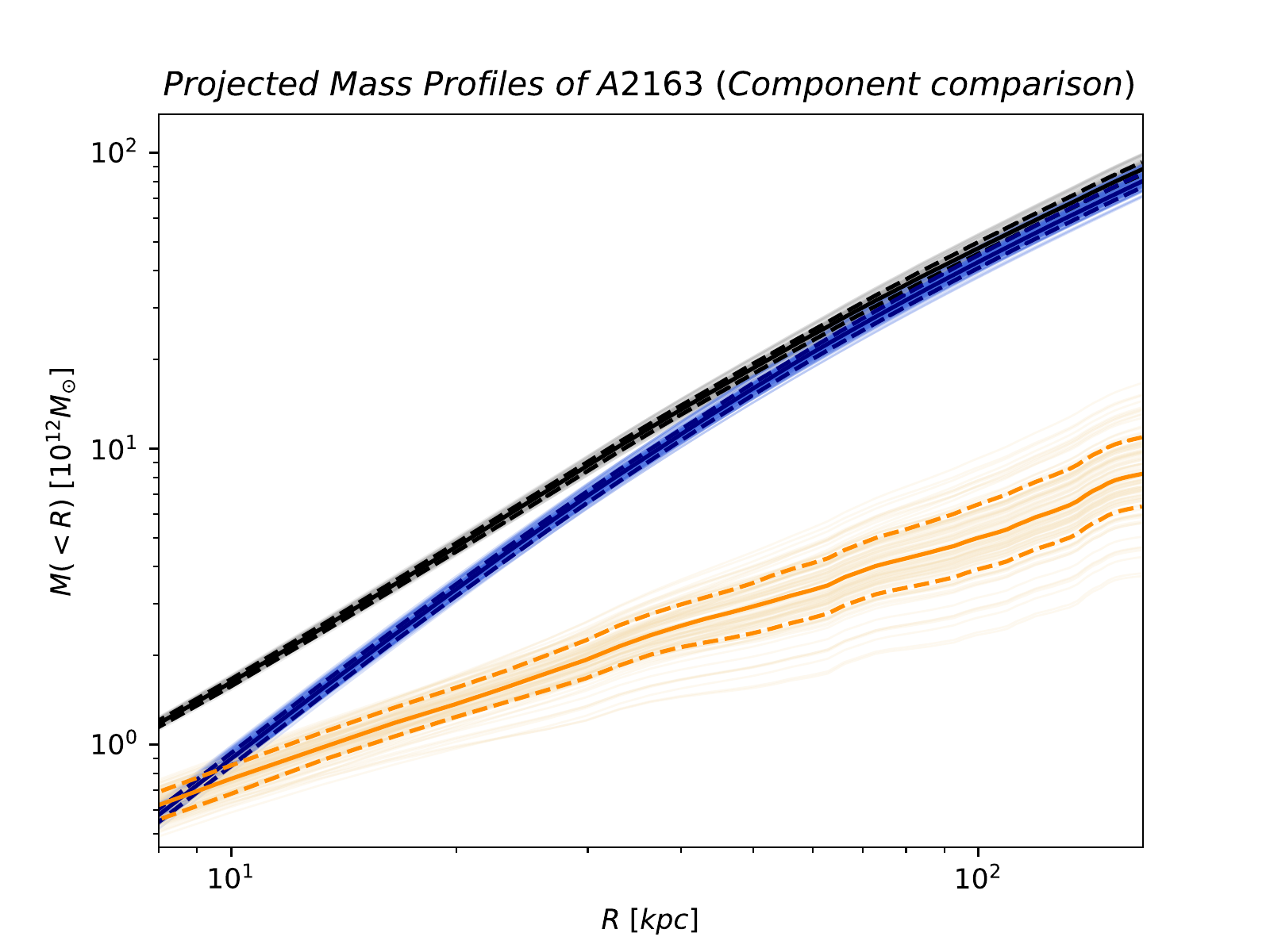}{\caption{\label{fig:Mass_Profile}Cumulative projected mass profiles from our fiducial model; black, blue and orange curves represent the total, the smooth and the clumpy components, respectively. Solid and dashed lines trace the median and 16th - 84th percentiles, while the light ones complete the sub-sample extracted from the final MCMC chains.}}
\end{figure}

\begin{figure*}[ht]
\centering
\includegraphics[width=\hsize]{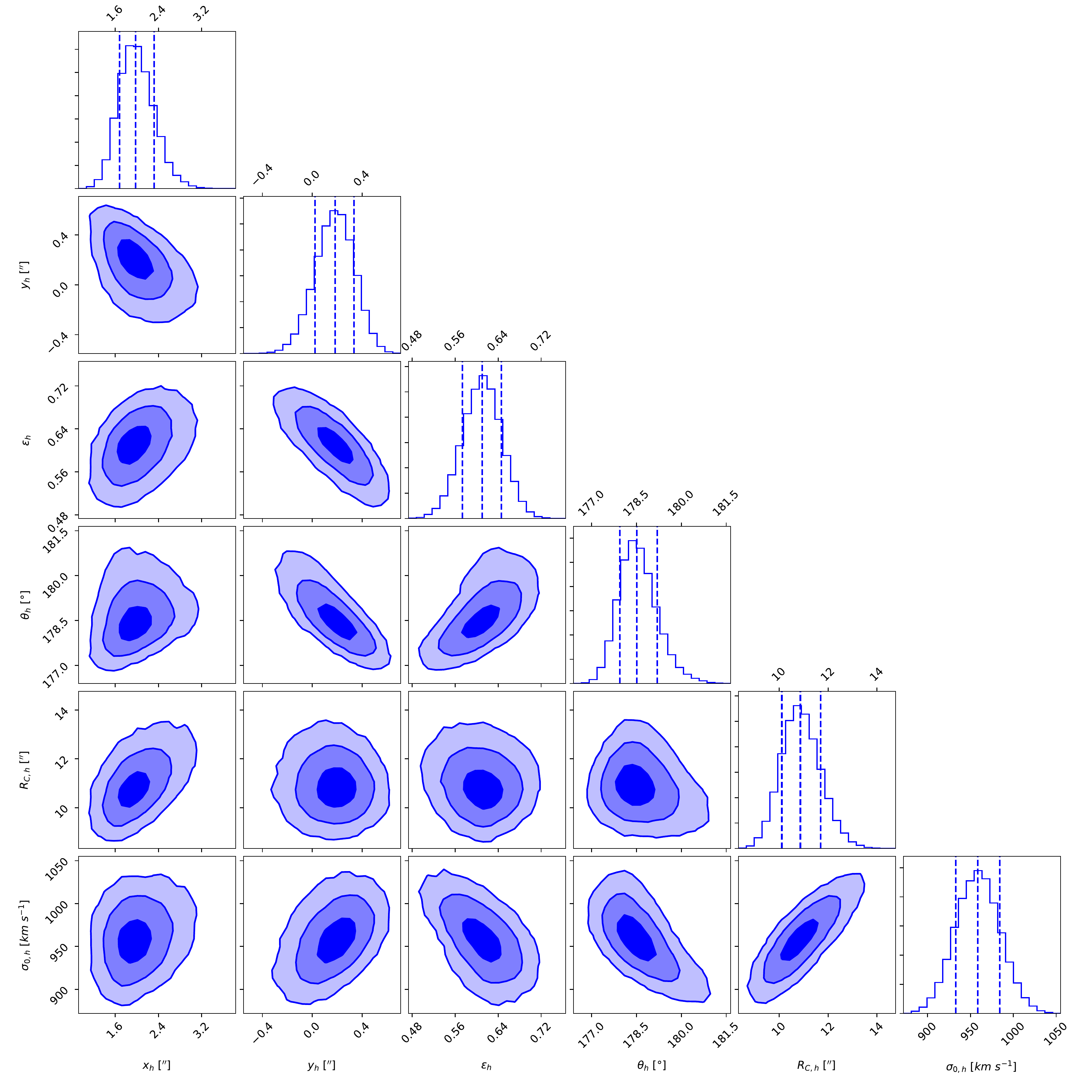}\caption{\label{fig:Pot1}Posterior distributions of the parameter values of the cluster-scale mass component. Blue contours correspond to the 1, 2 and 3 $\sigma$ confidence levels of a Gaussian distribution, while vertical blue dashed lines in the histograms are the 16th, 50th and 84th percentiles.}
\end{figure*}

\begin{figure}
\centering
\includegraphics[width=\hsize]{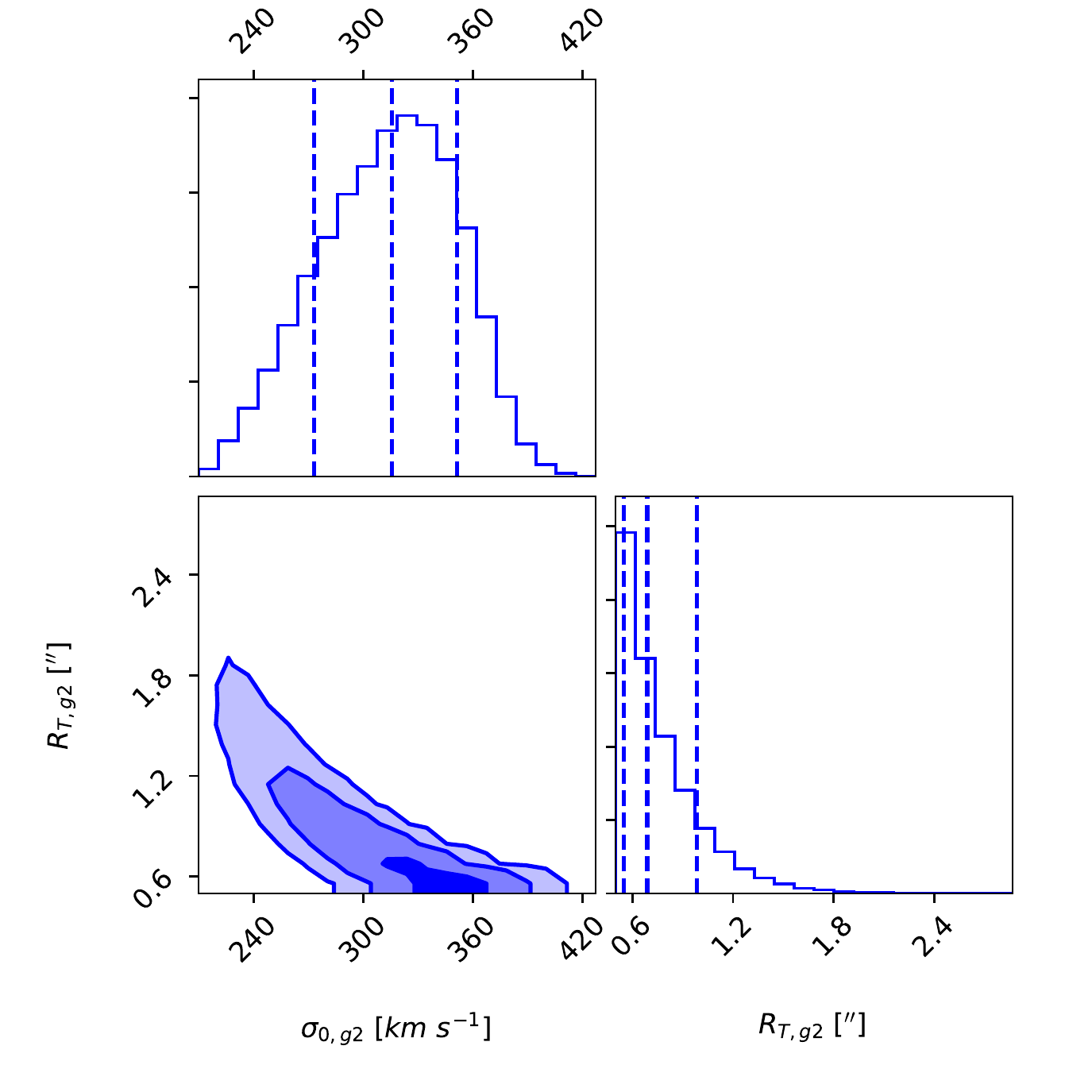}\caption{\label{fig:Pot2}As in Fig. \ref{fig:Pot1}, but for the two free parameters of the galaxy which contributes to the formation of F4 images.}
\end{figure}

\begin{figure}
\centering
\includegraphics[width=\hsize]{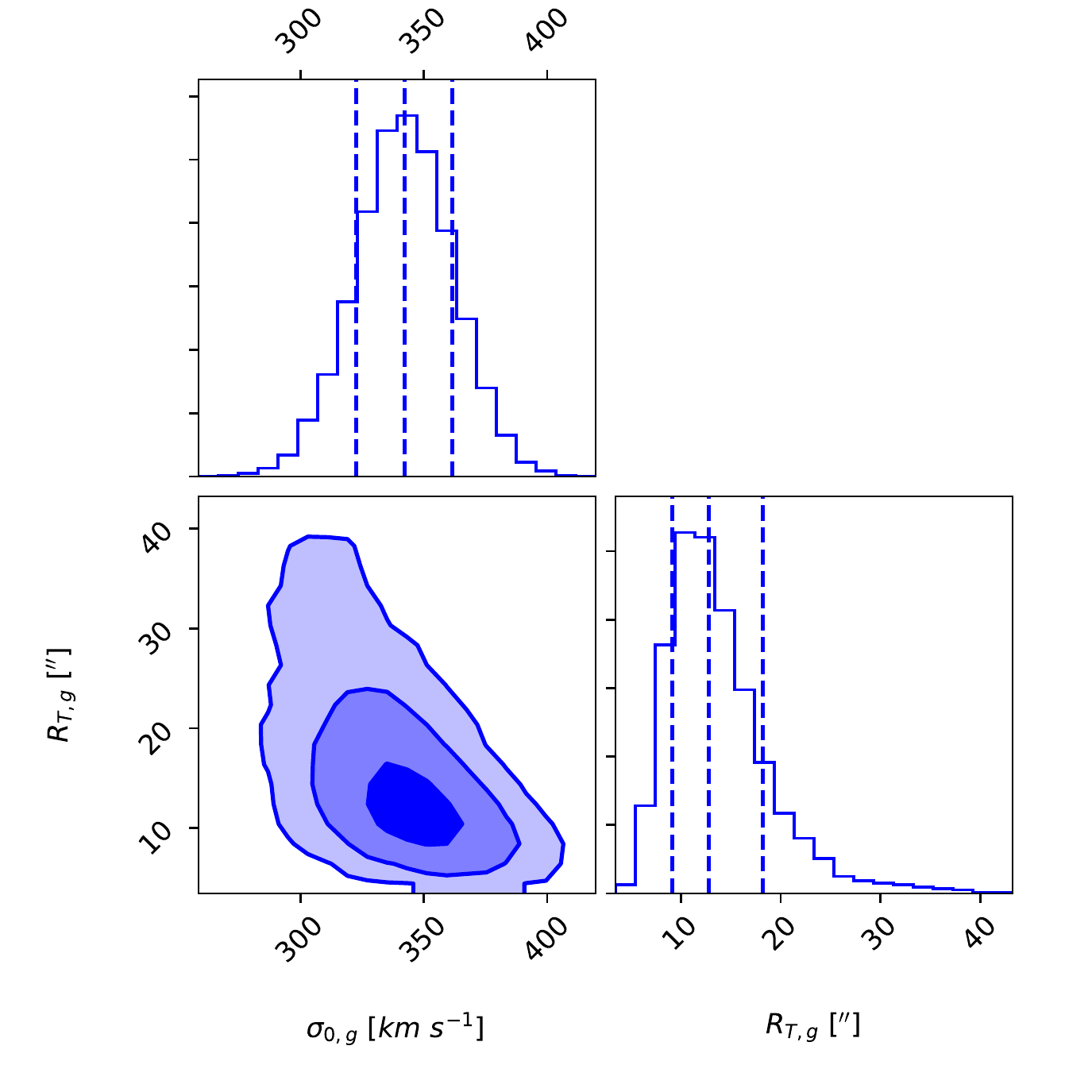}\caption{\label{fig:Pot0}As in Fig. \ref{fig:Pot1}, but for the BCG parameters, which are linked to the small-scale subhalo mass component, through the member galaxy scaling relations (\ref{eq:scaling_relation_R_T})-(\ref{eq:scaling_relation_sig_T}).}
\end{figure}

Our fiducial model predicts very accurately the positions of the multiple-image systems, with a final rms error $\delta_{\textnormal{rms}}\simeq0\arcsec.15$, i.e., approximately $2.5$ \textit{HST} pixels. Fig. \ref{fig:critical_lines} shows the critical curves corresponding to the three redshift values of the multiple image systems used in RUN 5.

We find a small (projected) distance of $\sim2\arcsec.0$ between the position of the diffuse component and the center of the BCG (see Fig. \ref{fig:Pot1}), with the last one towards east. Moreover, the smooth halo is flattened and elongated towards the A2163-A2 south-west direction.

The (median) values of the parameters and their errors can be found in Table \ref{tab:bf_params} and Figs. \ref{fig:Pot1}, \ref{fig:Pot2} and \ref{fig:Pot0}. Compared to the results by C18, we find smaller errors, but the parameter values are, overall, consistent. We remark, though, that the inclusion in the models of the spectroscopic redshift values for the strongly lensed sources alleviates the parameter value degeneracies, and thus significantly reduces the statistical error on the cumulative total mass profile of the cluster. In fact, our extrapolated value at $300$ kpc, $M(<300$ kpc$) = 1.43^{+0.07}_{-0.06}\times10^{14}\,\textnormal{M}_{\odot}$, is consistent, within the errors, with that found by C18, $M(<300$~kpc$) = {(1.6\pm{0.3})\times10^{14}\,\textnormal{M}_{\odot}}$.
\begin{table}
    \caption{Median values and confidence level (CL) uncertainties of the parameters for the lens model RUN 5 of A2163. Centers are relative to the BCG (R.A. = $243.9539405\si{\degree}$ and Decl. = $-6.1448406\si{\degree}$). Note that here the halo ellipticity is defined as $\epsilon_{\mathrm{h}}={(1-q^2)}/{(1+q^2)}$.
    }             
    \label{tab:bf_params} 
    \centering                  
    \begin{tabular*}{0.5\textwidth}{l c c c c}  
    \hline\hline                
    Parameter & Median & 68\% CL & 95\% CL & 99.7\% CL\\    
    \hline                    
   $x_{\mathrm{h}}$ [\arcsec] & 1.98 & $^{+0.34}_{-0.30}$ & $^{+0.74}_{-0.56}$ & $^{+1.19}_{-0.81}$	\\
   $y_{\mathrm{h}}$ [\arcsec] & 0.18 & $^{+0.15}_{-0.16}$ & $^{+0.29}_{-0.33}$ & $^{+0.42}_{-0.50}$	\\
   $\epsilon_{\mathrm{h}}$ & 0.61 & $^{+0.04}_{-0.04}$ & $^{+0.07}_{-0.07}$ & $^{+0.11}_{-0.11}$	\\
   $\theta_{\mathrm{h}}$ [$\si{\degree}$] & 178.5 & $^{+0.7}_{-0.6}$ & $^{+1.5}_{-1.1}$ & $^{+2.4}_{-1.6}$	\\
   $R_{\mathrm{C,h}}$ [kpc] & 36.0 & $^{+2.8}_{-2.5}$ & $^{+5.8}_{-4.8}$ & $^{+8.9}_{-6.8}$	\\
   $\sigma_{0,\mathrm{h}}$ [km s$^{-1}$] & 959 & $^{+26}_{-26}$ & $^{+52}_{-50}$ & $^{+77}_{-71}$	\\
    \hline                    
   $\sigma_{0,\mathrm{g2}}$ [km s$^{-1}$] & 316 & $^{+36}_{-43}$ & $^{+63}_{-79}$ & $^{+89}_{-97}$	\\
   $R_{\mathrm{T,g2}}$ [kpc] & 2.28 & $^{+0.98}_{-0.47}$ & $^{+2.53}_{-0.61}$ & $^{+4.93}_{-0.62}$	\\
    \hline                    
   $\sigma_{0,\mathrm{g}}$ [km s$^{-1}$] & 342 & $^{+20}_{-20}$ & $^{+39}_{-40}$ & $^{+58}_{-62}$	\\
   $R_{\mathrm{T,g}}$ [kpc] & 42.3 & $^{+18.0}_{-12.0}$ & $^{+50.5}_{-20.4}$ & $^{+87.5}_{-26.1}$	\\
    \hline                    
    \end{tabular*}
    \end{table}

From our new strong lensing model of A2163-A1, the cumulative projected total mass profile (relative to the position of the BCG) confirms that, in A2163-A1, the cluster halo is traced by its total light distribution. We also isolate the profiles for the two cluster mass components. The total, the halo, and the subhalo mass profiles are shown in Fig. \ref{fig:Mass_Profile}. There, three sets of curves are reported with different colors to distinguish between the total mass profile (in black) and those of the diffuse halo (in blue) and of the subhalos associated with cluster members (in orange). Solid and dashed lines identify the median and 16th $-$ 84th percentiles, respectively, while the light ones show a subsample in the final MCMC chains. The contribution to the total mass of the cluster-scale and the galaxy-scale components are $\sim90\%$ and $\sim10\%$, at $R>100$ kpc, respectively. Nevertheless, including mass substructures in a strong lensing model is fundamental to reconstructing a detailed cluster mass distribution for different reasons: firstly, to reproduce accurately the observed positions of the multiple images \citep{Kneib1996, Meneghetti2007, Meneghetti2017}, then, to understand the effective lens efficiency in the presence of a large numbers of perturbers and, finally, to avoid the introduction of systematic effects \citep{Jullo2007}.

\begin{figure}
\centering
\includegraphics[width=\hsize]{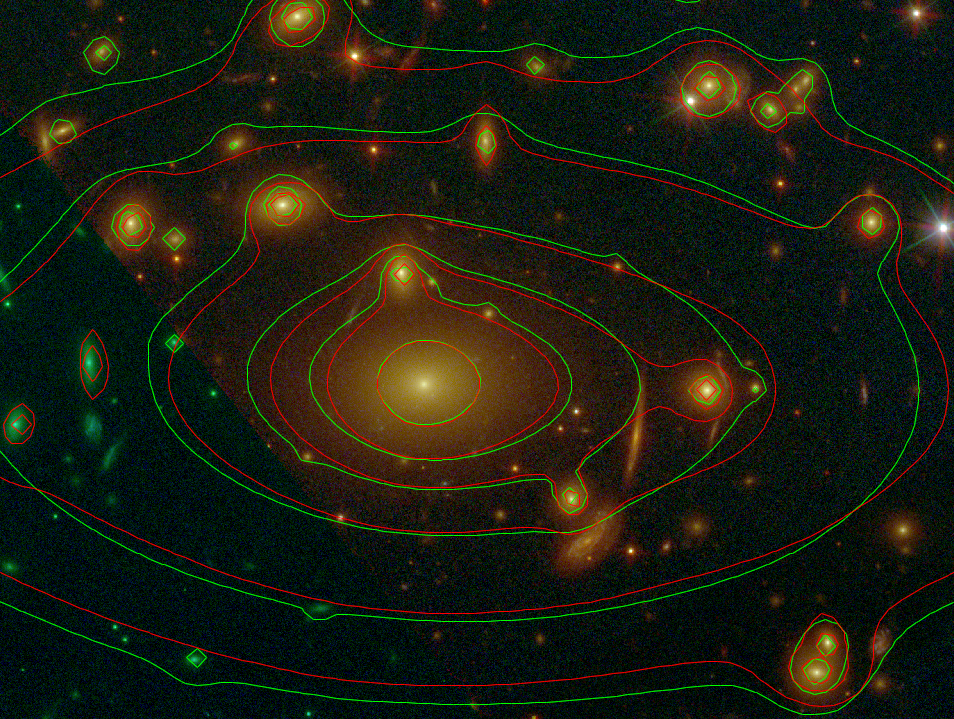}\caption{\label{fig:mass_map_comparison} Composite image of A2163-A1, with overlays of total surface mass density distribution. Green and red contours refer, respectively, to this work and C18. Contour line values are $[0.75\,,1.00\,,1.50\,,2.00\,,2.50\,,3.50]\times10^{9}\,\textnormal{M}_{\odot}\,\textnormal{kpc}^{-2}$.}
\end{figure}

Finally, a direct comparison with the total mass maps in the literature is not possible, since the needed information is not provided (e.g., in \citealt{Soucail2012}, the mass map contour levels derived from weak lensing analyses have no quoted values). The only exception is C18, whose convergence map is available on MAST. Figure \ref{fig:mass_map_comparison} shows the contour levels of the total surface mass density as derived in this work (in green) and in C18 (in red). We find an overall agreement, with only some variations which can be explained in terms of the different cluster member selection and subhalo total mass modeling. In particular, in this work, the values of the ellipticity and position angle of the subhalo component are not fixed to those obtained from the luminosity distribution of the cluster galaxies, because there is no conclusive evidence that the stellar mass elongation and orientation trace those of the total mass of a galaxy well at large distances from its center. Thus, we preferred to adopt simpler, circular total mass profiles to model the cluster subhalos. 

\section{Conclusions}\label{sec:Conclusions}

In this paper, we concentrate on the innermost region of the cluster A2163. We present our redshift measurements for all the member galaxies in the cluster core and within our MUSE FoV, compiling a pure spectroscopic catalog. Then, we extend it to a subsample of member candidates selected on the basis of their photometric information. We also report the discovery of new multiple images and, exploiting our novel MUSE data, spectroscopically confirm the majority of those already found in the literature. With such a solid dataset, we built a new strong lensing model, using \textit{HST} positions of multiple-image systems as constraints, and finally, we determined an accurate projected total mass profile in the core of A2163. We also decomposed the profile in the two cluster- and galaxy-scale components, clarifying which parameters of the diffuse mass distribution are in favor of a possible cluster merging scenario.

The main results can be summarized as follows:
\begin{enumerate}
\item{We have measured the spectra of more than 200 sources in the A2163 core and spectroscopically confirmed 35 foreground and background sources (see the table of Appendix \ref{Appendix_A}), 29 cluster members (Table \ref{tab:spec_members_cat}) and 8 multiple images.}
\item{We have discovered 4 new multiple images and presented a final catalog consisting of 16 multiple images from 4 different sources, with redshifts from 1.16 up to 2.72 (Table \ref{tab:images_cat}).}
\item{We have compared the predicted positions of the multiple images with those of their observational counterparts and found, for our fiducial model, an rms value of $\delta_{\textnormal{rms}}=0\arcsec.15$.}
\item{After testing different lensing models, we have concluded that the projected total mass distribution of A2163-A1 can be well represented by a diffuse component with a dPIE profile, a galaxy-scale spherical halo and a population of 110 sub-halos, with a total M/L ratio increasing with luminosity, as observed in early-type galaxies (tilt of the fundamental plane).}
\item{The shape of the diffuse halo of A2163-A1, elongated in the direction of A2163-A2, supports a scenario in which, after the merger, the halo of the former has relaxed to an elliptical shape which points to the latter. In fact, although the halo center has been found within $\sim2\arcsec$ from the BCG, the X-ray main peak (in \citealt{Maurogordato2008}) is far from being coincident with their (nearly common) positions, which disfavors a pre-merging phase. We note that our estimated values of the halo ellipticity and position angle have an error about an order of magnitude smaller than previously found. On larger scales, a confirmation of this merging scenario is not possible with strong lensing only.}
\item{From our fiducial model, we have measured the cluster cumulative projected total mass profile very precisely and have found that $M(<100$ kpc$) = 4.75^{+0.23}_{-0.20}\times10^{13}\,\textnormal{M}_{\odot}$, which is consistent with the value with significantly larger statistical errors found in the literature.}
\item{When a diffuse mass component (mostly of DM) is used to model A2163 core, its center is found between the BCG and the main X-ray peak, identified by \cite{Maurogordato2008} between the east and the west sub-structures in the cluster core. Models which separate the mass contributions of the galaxies, the DM and the hot gas need to be explored, as they might mitigate the offset between the BCG and the diffuse halo. In this direction, a relevant and new approach has been presented for the cluster MACS J0416.1-2403: in \cite{Bonamigo2017}, the total mass and hot gas distributions have been separated and, in \cite{Annunziatella2017}, the further subtraction of the stellar component and the consequent decoupling of the DM distribution has led to a complete mass decomposition. If these models will confirm a significant difference between the centers of the stellar and DM components, we should not exclude the self-interaction of DM as a possible explanation for it. In fact, the relative positions and the alignment of the mass components of our fiducial model are consistent with the predictions of such a scenario for merging sub-clusters.} 
\end{enumerate}

The authors thank the anonymous referee for the constructive comments that helped to improve the quality of this paper. U. R. warmly thanks Keren Sharon for the useful suggestions on the manuscript and Sebasti{\'a}n L{\'o}pez for his kind hospitality at the \textit{Universidad de Chile}, where part of this work was done. A. M. acknowledges funding from the INAF PRIN-SKA 2017 program 1.05.01.88.04. This work is based on observations taken by the RELICS Treasury Program (GO 14096) with the NASA/ESA HST, which is operated by the Association of Universities for Research in Astronomy, Inc., under NASA contract NAS5-26555. Corner plots were created using the corner.py module by \cite{Foreman-Mackey2016}.

\bibliographystyle{aa} 
\bibliography{MyBibliography} 

\Online

\onecolumn

\begin{appendix} 
\section{catalog of redshift measurements of foreground and background sources.}\label{Appendix_A}
In the following catalog, we report \textit{HST} IDs, celestial coordinates, and spectroscopic redshifts (with QFs) for all sources other than cluster galaxies and multiple images, whose redshift measurements are, respectively, in Table \ref{tab:spec_members_cat} and Table \ref{tab:images_cat}. We exclude sources with QF $<1$, because they are too faint and/or too noisy and do not show clear spectroscopic features.

\begin{longtab}
\begin{longtable}{c c c c c}
\hline\hline                      
ID & R.A. [Deg] & Decl. [Deg] & $z_{\mathrm{sp}}$ & QF \\    
\hline                            
3192	 & 	243.94458	 & 	$-$6.13500	 & 	0.0	 & 	3	 \\
3710	 & 	243.94535	 & 	$-$6.14001	 & 	0.0	 & 	3	 \\
3356	 & 	243.94752	 & 	$-$6.13804	 & 	0.0	 & 	3	 \\
4430	 & 	243.94896	 & 	$-$6.14884	 & 	0.0	 & 	3	 \\
3818	 & 	243.95029	 & 	$-$6.14547	 & 	0.0	 & 	3	 \\
3819	 & 	243.95066	 & 	$-$6.14589	 & 	0.0	 & 	3	 \\
3817	 & 	243.95176	 & 	$-$6.14685	 & 	0.0	 & 	3	 \\
3594	 & 	243.95516	 & 	$-$6.13921	 & 	0.0	 & 	3	 \\
3391	 & 	243.95563	 & 	$-$6.13696	 & 	0.0	 & 	3	 \\
4545	 & 	243.95596	 & 	$-$6.14804	 & 	0.0	 & 	3	 \\
3431	 & 	243.95760	 & 	$-$6.13750	 & 	0.0	 & 	3	 \\
4219	 & 	243.95904	 & 	$-$6.14505	 & 	0.0	 & 	3	 \\
3902	 & 	243.95992	 & 	$-$6.14182	 & 	0.0	 & 	3	 \\
3802	 & 	243.96003	 & 	$-$6.13957	 & 	0.0	 & 	3	 \\
3886	 & 	243.96081	 & 	$-$6.14225	 & 	0.0	 & 	3	 \\
4791	 & 	243.96117	 & 	$-$6.15096	 & 	0.0	 & 	3	 \\
5256	 & 	243.96141	 & 	$-$6.15066	 & 	0.0	 & 	3	 \\
3240	 & 	243.95248	 & 	$-$6.13561	 & 	0.0	 & 	2	 \\
\hline                                   
3726	 & 	243.95371	 & 	$-$6.14011	 & 	0.3344	 & 	3	 \\
3371	 & 	243.95047	 & 	$-$6.13713	 & 	0.3728	 & 	1	 \\
3868	 & 	243.95238	 & 	$-$6.14197	 & 	0.4075	 & 	3	 \\
3242	 & 	243.95538	 & 	$-$6.13565	 & 	0.4399	 & 	3	 \\
3576	 & 	243.94649	 & 	$-$6.13783	 & 	0.5940	 & 	3	 \\
4338	 & 	243.96156	 & 	$-$6.14658	 & 	0.6640	 & 	3	 \\
4147	 & 	243.94883	 & 	$-$6.14609	 & 	0.6646	 & 	3	 \\
4152	 & 	243.94896	 & 	$-$6.14678	 & 	0.6657	 & 	3	 \\
4278	 & 	243.96196	 & 	$-$6.14585	 & 	0.6658	 & 	3	 \\
4463	 & 	243.95000	 & 	$-$6.14856	 & 	0.7284	 & 	3	 \\
9000\footnote{This ID is assigned to a source which we identify and that is not present in the \textit{HST} catalog.}	 & 	243.95826	 & 	$-$6.14134	 & 	0.8249	 & 	3	 \\
4492	 & 	243.96150	 & 	$-$6.14761	 & 	0.8673	 & 	1	 \\
4644	 & 	243.94810	 & 	$-$6.14873	 & 	0.9787	 & 	3	 \\
3284	 & 	243.96128	 & 	$-$6.13606	 & 	1.0406	 & 	1	 \\
3568	 & 	243.94898	 & 	$-$6.13870	 & 	1.1664	 & 	3	 \\
3580	 & 	243.95434	 & 	$-$6.13708	 & 	1.3725	 & 	3	 \\
3285	 & 	243.95490	 & 	$-$6.13686	 & 	1.3740	 & 	3	 \\
3222	 & 	243.95373	 & 	$-$6.13571	 & 	1.3775	 & 	3	 \\
4713	 & 	243.95345	 & 	$-$6.14944	 & 	3.1660	 & 	1	 \\
3854	 & 	243.94453	 & 	$-$6.14187	 & 	4.5837	 & 	2	 \\
3693	 & 	243.94649	 & 	$-$6.13971	 & 	4.9901	 & 	2	 \\
\hline                                   
\end{longtable}
\end{longtab}

\end{appendix}

\end{document}